# Dual chain perturbation theory: A new equation of state for polyatomic molecules

Bennett D. Marshall

*ExxonMobil Research and Engineering, 22777 Springwoods Village Parkway, Spring TX 77389*

## Abstract

In the development of equations of state for polyatomic molecules, thermodynamic perturbation theory (TPT) is widely used to calculate the change in free energy due to chain formation. TPT is a simplification of a more general and exact multi – density cluster expansion for associating fluids. In TPT all contributions to the cluster expansion which contain chain – chain interactions are neglected. That is, all inter-chain interactions are treated at the reference fluid level. This allows for the summation of the cluster theory in terms of reference system correlation functions only. The resulting theory has been shown to be accurate, and has been widely employed as the basis of many engineering equations of state. While highly successful, TPT has many handicaps which result from the neglect of chain – chain contributions. The subject of this document is to move beyond the limitations of TPT, and include chain – chain contributions to the equation of state.

bennett.d.marshall@exxonmobil.com

**I: Introduction**

Polyatomic molecules consisting of chains of covalently bonded atoms are common both in nature and industry. The development of equations of state for chain molecules is made difficult by the anisotropy of intermolecular interactions (non – spherically symmetric). In a simplified picture of chain molecules, chains can be modelled as chains of tangentially bonded spheres. Chandler and Pratt (CP) were the first to develop a theoretical formalism to describe flexible chains of spheres[1]. The contribution of CP was to extend the cluster theory of Morita and Hiroike[2] to the case of polyatomic molecules with internal degrees of freedom. CP grouped monomers into physical clusters bonded by covalent bonds, and then topologically reduced the cluster expansion from a fugacity basis to a density basis. This topological reduction to density graphs "normalized" the theory, eliminating any reference to the infinitely large bonding Mayer functions.

Nearly a decade later, Wertheim[3,4] developed a new cluster expansion for associating molecules. Wertheim's key breakthrough was to use a multi-density approach, where each bonding state of a molecule had its own assigned density. Topological reduction from fugacity graphs to these multi – density graphs, renormalized the theory for the case of strong covalent bonds by pairing infinitely large Mayer functions with infinitely small monomer densities. Chapman[5] was the first to show, that when the multi – density approach as applied to the case of very strong association (covalent bonding), the association theory could be used to develop equations of state for chain molecules. The change in free energy due to chain formation was derived in thermodynamic perturbation theory (TPT) which, in the context of Wertheim's multi-density approach, means all contributions to the free energy were retained which accounted for interactions of the monomer reference fluids, and interactions of a single chain with the reference fluid. However, all contributions to the free energy which contain inter-chain interactions are neglected. This is also known as the single chain approximation, and it allows for the description of the properties of the chain fluid in terms of reference system properties only.[6] In short, TPT, corrects for intra-chain effects, but neglects inter-chain corrections to the reference fluid free energy.

TPT has been a very useful tool in the description of chain molecules fluids, evidenced by the number of engineering equations of state[7–10] which employ this approach. Since its inception, there have been several advances to the TPT equation of state[11–13]; however, there has been no attempt to extend the chain equation of state beyond TPT. There are a number of situations where this would be desirable. For instance, exact predictions of the second virial coefficient, effects of branching on intermolecular interactions and liquid crystals all require the accounting of chain – chain interactions, which are completely neglected in TPT. One approach is to employ the more complicated formalism of integral equation theory[14,15]. Alternatively, one could develop a new TPT structure which includes the hierarchy of diagrams which contain two or more interacting chains. In this work we take the latter approach and develop a correction to TPT which accounts for the interaction of two chain molecules in the presence of the reference fluid.

**II: Theory**

In this section we develop a new free energy for chain molecules in Wertheim's two density formalism.[16] The theory derived here, could similarly be derived in Wertheim's multi – density formalism[4] or Chandler and Pratt (CP)[1] density functional theory. The two density approach assigns molecules separate densities depending on their bonding state. Molecules which are bonded receive a density $\rho_b(1)$ and molecules which are not bonded (monomers) receive a density $\rho_o(1)$. The total density of molecules in the fluid is then given by $\rho(1) = \rho_o(1) + \rho_b(1)$. The notation $(1) = \{\vec{r}_1, \Omega_1\}$ represents the position and orientation of molecule 1. Wertheim's two density formalism is typically reserved for associating fluids with a single association site, with multi – site fluids and chain formation reserved for the multi – density (>2) approach. However, there is no fundamental reason why the two density approach cannot be applied to chain formation. In fact, for chain formation, the two density approach is equivalent to the multi – density approach. The advantage of the two density approach is simplified notation and a more transparent derivation. In regards to CP theory, which is well suited for the derivation of chain equations of state, we choose Wertheim's approach due to the easy extension of these results to associating fluids, which are more conveniently derived in Wertheim's theory. That said, the approach presented here will have many features of CP. In fact, the approach will be a hybrid of Wertheim's two density formalism and CP density functional theory.

### A: Associating fluids

In the development of the theory of chain molecules we will initially use the language of associating fluids. Specifically, we consider a fluid of associating spheres which interact with the following intermolecular potential

$$\phi(12) = \phi_R(12) + \phi_A(12) \tag{1}$$

Equation (1) splits the intermolecular potential into reference $\phi_R$ and association $\phi_A$ contributions. In the two density formalism the Helmholtz free energy is given exactly as[16]

$$\frac{A}{k_B T} = \int \left( \rho(1) \ln(\rho_o(1) \Lambda^3) - \rho_o(1) \right) d(1) - c^{(o)} \tag{2}$$

The term $c^{(o)}$ is an infinite series of integrals given by

$c^{(o)}$ = *sum of all irreducible diagrams consisting of monomer points carrying* (3)
*factors of $\rho$, m-mer subgraphs with m $\geq$ 2 and every point carrying a*
*factor of $\rho_o$ and $f_R$ bonds between some points in distinct m-mers*

In Eq (3), an *m*-mer subgraph is a graph which contains *m* points connected by association bonds $F_A(12)$. Equation (4) defines the types of Mayer functions used in this work

$$e_R(12) = \exp\left(-\frac{\phi_R(12)}{k_B T}\right) \qquad f_R(12) = e_R(12) - 1$$

$$f_A(12) = \exp\left(-\frac{\phi_A(12)}{k_B T}\right) - 1 \qquad F_A(12) = e_R(12) f_A(12)$$

(4)

Figure 1 illustrates the first few terms in the sum Eq. (3) for the case of a dimerizing system. In this work we restrict attention to the diagrams with only a single type of "associated" cluster, or *m*-mer. For ease of presentation, we will initially restrict our attention to linear chains. However, the derived results will be easily extended to other classes of molecules such a branched chains and rings. We only consider association in chains of length *m*, disregarding chains with lengths smaller than *m* which would necessarily be created in the process. However, our main goal is irreversible chain formation. Once we take the limit of complete association, these shorter chain contributions will be negligible.

The sum $c^{(o)}$ is now decomposed as follows

$$c^{(o)} = c_o^{(o)} + c_1^{(o)} + c_2^{(o)} + c_3^{(o)} + \cdots + c_\infty^{(o)}$$ (5)

Where $c_k^{(o)}$ represents the infinite sum diagrams with $k$ $m$-mer chains interacting with themselves and the reference fluid. Note, this *is not* the thermodynamic perturbation theory (TPT) expansion used by Wertheim[6] and others[17]. The contribution $k = 1$ alone gives TPT to infinite order. The contribution for $k = 0$ gives all diagrams for the reference fluid. In Fig. 1, the reference fluid diagrams are given by **a**, **c**, **g**, **h** and **i**. The sum of all reference fluid diagrams is given simply as

$$c_o^{(o)} = -\frac{A_R^{EX}}{k_B T} \tag{6}$$

where $A_R^{EX}$ is the excess free energy of the reference fluid. The contribution for $k = 1$ gives all diagrams representing the interaction of a single chain with the reference fluid. In Fig. 1 the diagrams **b**, **d**, **j**, **k**, **l** and **m** belong to this class. This infinite sum may then be condensed in terms of the $m$ – body reference system correlation function as

$$c_1^{(o)} = \frac{1}{2}\int g_R(1_m)f_A(12)\cdots f_A(m-1,m)\prod_{k=1}^{m}\rho_o(k)d(k) = \frac{1}{2}\int s_R(1_m)d(1_m) \tag{7}$$

Where we have introduced the intramolecular distribution functions

$$s_R(1_m) = g_R(1_m)f_A(12)\cdots f_A(m-1,m)\prod_{k=1}^{m}\rho_o(k) \tag{8}$$

These are related to the contracted intramolecular distribution functions of CP. We have also adapted the notation similar to CP notation, where $1^{(k)}$ represents the position and orientation of molecule $k$ in associated cluster 1 and $1_m = \{1^{(1)}\cdots m^{(1)}\}$ represents the position and orientation of all molecules in the cluster. Finally, $d(1_m)$ represents integration over all position and orientations of all molecules in the cluster.

All previous chain equations[12,13,18] of state derived from Wertheim's multi-density formalism have truncated the expansion of Eq. (5) at $k = 1$. This is known as the single chain approximation[6] and forms the basis of thermodynamic perturbation theory. At this level of approximation the theory corrects for the *intra*molecular contributions of chain formation, but does not address the changes in *inter*molecular interactions which result from forming chains from a fluid of spheres. This introduces a number of limitations in the theory such as the inability to predict the exact second virial coefficient of chain molecules, liquid crystals, effect of branching on intermolecular interactions etc.... To include this information we must move beyond the single chain approximation. Meaning, at a minimum, we must include some subset of diagrams from $k = 2$. Diagrams **n** and **o** in Fig. 1 belong to this class. To this end we consider the following subset of diagrams in the $k = 2$ sum

$c_2^{(o)} =$ *The sum of diagrams in Eq. (3) which contain two m-mers and any number* (9)
  *of monomer points where the only path between the two m-mers are direct $f_R$*
  *bonds between segments in the two m-mers*

The sum given in Eq. (9) will account for the direct interactions between 2 chain molecules in the presence of the reference fluid, but does not account for indirect correlation between two chain molecules where, for instance, two chain molecules are side by side with a thin solvent layer between.

The restriction in Eq. (9) that the only path of bonds between *m*-mers is direct $f_R$ bonds allows the sum $c_2^{(o)}$ to be topologically reduced to the following

$$c_2^{(o)} = \frac{1}{4}\int s(1_m)s(2_m)\hat{\Gamma}(1_m,2_m)d(1_m)d(2_m) \tag{10}$$

Where the density independent function $\hat{\Gamma}(1_m,2_m)$ is obtained by taking all of the ways to connect the two *m*-mers with $f_R$ bonds such that the diagrams in Eq. (9) are irreducible. For the case $m = 2$ we find

$$\hat{\Gamma}(1_m,2_m) = f_R(1^{(1)},2^{(1)})f_R(1^{(2)},2^{(2)}) + 2f_R(1^{(1)},2^{(1)})f_R(1^{(2)},2^{(2)})f_R(1^{(1)},2^{(2)}) \qquad (11)$$

$$+ \frac{1}{2} f_R(1^{(1)},2^{(1)})f_R(1^{(2)},2^{(2)})f_R(1^{(1)},2^{(2)})f_R(1^{(2)},2^{(1)})$$

Before proceeding further, we introduce cavity correlation functions $y_R$ defined by

$$g_R(1_m) = y_R(1_m)E_R(1_m) \quad ; \quad E_R(1_m) = \prod_{k=2}^{m}\prod_{l=1}^{k-1} e_R(lk) \qquad (12)$$

Now we normalize the association Mayer functions

$$f_A(12) = \lambda \hat{f}_A(12) \qquad (13)$$

Where the normalized association Mayer functions $\hat{f}_A$ are defined by Eq. (13) and $\lambda$ is the average association strength

$$\lambda = \frac{\int f_A(12)d(2)}{\Delta} \qquad (14)$$

$\Delta$ in Eq. (14) gives the association "bond volume".

From this point forward we assume a homogeneous fluid

$$\rho_o = \int d\Omega_1 \rho_o(1) = \rho_o(1)\Omega \qquad (15)$$

where $\Omega$ is the orientation volume (dimensionless). Using (12) – (13) and (15) we simplify Eq. (8) as

$$s_R(1_m) = \frac{1}{\Omega^m}\rho_o^m \lambda^{m-1} y_R(1_m)E_R(1_m)\prod_{k=1}^{m-1}\hat{f}_A(1^{(k)},1^{(k+1)}) = \frac{1}{\Omega}\rho_o^m \lambda^{m-1} y_R(1_m) b^{\text{intra}}(1_m) \qquad (16)$$

The term $b^{\text{intra}}$ contains all density independent intra-chain contributions

$$b^{\text{intra}}(1_m) = E_R(1_m)\prod_{k=1}^{m-1}\frac{\hat{f}_A(1^{(k)},1^{(k+1)})}{\Omega} \qquad (17)$$

Using Eq. (16) we simplify Eq. (7)

$$\frac{c_1^{(o)}}{V} = \frac{\rho_o^m \lambda^{m-1}}{2V\Omega}\int y_R(1_m)b^{\text{intra}}(1_m)d(1_m) = \frac{\rho_o^m \lambda^{m-1}}{2}\langle y_R(1_m)\rangle_s \frac{\int b^{\text{intra}}(1_m)d(1_m)}{V\Omega} \qquad (18)$$

The angular brackets represent an average over the associated states of the cluster

$$\langle y_R(1_m) \rangle_s = \frac{\int b^{\text{intra}}(1_m) y_R(1_m) d(1_m)}{\int b^{\text{intra}}(1_m) d(1_m)} \tag{19}$$

Similiarly for the $c_2^{(o)}$ contribution

$$\frac{c_2^{(o)}}{V} = \left(\frac{\rho_o^m \lambda^{m-1}}{2}\right)^2 \langle y_R(1_m) y_R(2_m) \rangle_d \frac{\int b^{\text{intra}}(1_m) b^{\text{intra}}(2_m) \hat{\Gamma}(1_m, 2_m) d(1_m) d(2_m)}{V \Omega^2} \tag{20}$$

where the brackets $\langle \; \rangle_d$ represents the average over the configurations of two chains and the interaction between segments on the chains

$$\langle y_R(1_m) y_R(2_m) \rangle_d = \frac{\int b^{\text{intra}}(1_m) b^{\text{intra}}(2_m) \hat{\Gamma}(1_m, 2_m) y_R(1_m) y_R(2_m) d(1_m) d(2_m)}{\int b^{\text{intra}}(1_m) b^{\text{intra}}(2_m) \hat{\Gamma}(1_m, 2_m) d(1_m) d(2_m)} \tag{21}$$

Before proceeding further, we characterize the sum $\hat{\Gamma}(1_m, 2_m)$ and the integral in Eq. (20) in terms of the second virial coefficient of chain molecules. Written in terms of the $1_m$ notation, we can recast the second virial coefficient of chain molecules as

$$B_{2c} = -\frac{1}{2V\Omega^2 d^{6(m-1)} \Xi_m^2} \int b^{\text{intra}}(1_m) b^{\text{intra}}(2_m) \left( \exp\left(-\frac{1}{k_B T} \sum_{k=1}^{m} \sum_{l=1}^{m} \phi_R(1^{(k)}, 2^{(l)})\right) - 1 \right) d(1_m) d(2_m) \tag{22}$$

Where $\Xi_m$ is the isolated chain partition function

$$\Xi_m = \frac{\int b^{\text{intra}}(1_m) d(1_m)}{d^{3(m-1)} \Omega V} \tag{23}$$

Expanding the exponential in Eq. (22) in terms of Mayer functions between segments

$$\frac{1}{2}\left[\exp\left(-\frac{1}{k_B T} \sum_{k=1}^{m} \sum_{l=1}^{m} \phi_R(1^{(k)}, 2^{(l)})\right) - 1\right] = \frac{1}{2} \prod_{k=1}^{m} \prod_{l=1}^{m} [1 + f_R(1^{(k)}, 2^{(l)})] - \frac{1}{2} \tag{24}$$

$$= \hat{\Gamma}(1_m, 2_m) + \hat{C}(1_m, 2_m) + \hat{D}(1_m, 2_m)$$

The sum $\hat{C}(1_m, 2_m)$ contains all contributions which involve two or more $f_R(1^{(k)}, 2^{(l)})$, but only one segment on one of the chains has an incident $f_R$ bond. Now comparing to Eq. (9), the $\hat{C}(1_m, 2_m)$ contribution is absent due to the fact this connectivity gives rise to reducible diagrams.

The sum $\hat{D}(1_m, 2_m)$ contains all contributions in which there is only a single $f_R$ bond between the two $m$ – mers. Again, these contributions are absent in (9) because this connectivity gives rise to reducible diagrams. Combining (22) and (24)

$$B_{2c} = -\frac{1}{V\Omega^2 \Xi_m^2 d^{6(m-1)}} \int b^{\text{intra}}(1_m) b^{\text{intra}}(2_m) \left( \hat{\Gamma}(1_m, 2_m) + \hat{C}(1_m, 2_m) + \hat{D}(1_m, 2_m) \right) d(1_m) d(2_m) \quad (25)$$

Using Eq. (25) to eliminate the integral in Eq. (20)

$$\frac{c_2^{(o)}}{V} = \left( \frac{d^{3(m-1)} \rho_o^m \lambda^{m-1}}{2} \right)^2 \Xi_m^2 \langle y_R(1_m) y_R(2_m) \rangle_d \Gamma \quad (26)$$

Where

$$\Gamma = C + D - B_{2c} \quad (27)$$

With $D$ given by the integral

$$D = -\frac{1}{V\Omega^2 \Xi_m^2 d^{6(m-1)}} \int b^{\text{intra}}(1_m) b^{\text{intra}}(2_m) \hat{D}(1_m, 2_m) d(1_m) d(2_m) \quad (28)$$

Equation (28) can be evaluated in terms of the second virial coefficient of the reference system $B_R$ as

$$\frac{D}{m^2} = -2\pi \int f_R(r) r^2 dr = B_R \quad (29)$$

Similarly, $C$ in Eq. (27) is given by

$$C = -\frac{1}{V\Omega^2 \Xi_m^2 d^{6(m-1)}} \int b^{\text{intra}}(1_m) b^{\text{intra}}(2_m) \hat{C}(1_m, 2_m) d(1_m) d(2_m) \quad (30)$$

As will be demonstrated shortly, $C$ can be related to the low density limit of the density derivative of the correlation function $y_R(1_m)$.

With (26) we have completely defined $c^{(o)}$ for the association of $m$ molecules into a chain of size $m$. Minimizing the free energy Eq. (2) with respect to monomer density, gives the following closed relation for monomer densities

$$\rho = \rho_o + \frac{md^{3(m-1)}\rho_o^m \lambda^{m-1}}{2}\langle y_R(1_m)\rangle_s \Xi_m + 2m\left(\frac{d^{3(m-1)}\rho_o^m \lambda^{m-1}}{2}\right)^2 \Xi_m^2 \langle y_R(1_m)y_R(2_m)\rangle_d \Gamma \tag{31}$$

Equation (31) is generally valid for the association of $m$ molecules into chains of size $m$ only. To develop an equation of state for hydrogen bonding fluids, one would need to include contributions to $c^{(o)}$ for chains of all sizes $m = 2 - \infty$. The goal here is to develop the change in free energy due to formation of chain molecules where the segments in the chain are irreversible bonded.

**B: Chain formation**

To irreversibly bond spherical segments into chains, we let the association strength become infinitely large $\lambda \to \infty$ which drives the monomer density to zero $\rho_o \to 0$. This allows for the neglect of the $\rho_o$ on the right hand side of (31), and the solution of this equation as

$$d^{3(m-1)}\rho_o^m \lambda^{m-1} = \frac{1}{\Xi_m \Gamma} \frac{\langle y_R(1_m)\rangle_s}{\langle y_R(1_m)y_R(2_m)\rangle_d} \frac{\xi - 1}{2} \tag{32}$$

where

$$\xi = \sqrt{1 + 8\rho_c \Gamma \frac{\langle y_R(1_m)y_R(2_m)\rangle_d}{\langle y_R(1_m)\rangle_s^2}} \tag{33}$$

and $\rho_c = \rho/m$ is the density of chain molecules. Using Eq. (32) to eliminate the monomer density in Eq. (18)

$$\frac{c_1^{(o)}}{V} = \frac{\langle y_R(1_m)\rangle_s^2}{\langle y_R(1_m)y_R(2_m)\rangle_d} \frac{\xi-1}{4\Gamma} \tag{34}$$

and Eq. (26)

$$\frac{c_2^{(o)}}{V} = \frac{\langle y_R(1_m)\rangle_s^2}{\langle y_R(1_m)y_R(2_m)\rangle_d} \frac{(\xi-1)^2}{16\Gamma} \tag{35}$$

Now, to simplify the Helmholtz free energy in Eq. (2), we assume a homogeneous fluid in the limit of complete association, with $c^{(o)}$ obtained from (34) and (35) and the monomer density within the natural log eliminated using Eq. (32) to obtain

$$\frac{A - A_R^{EX}}{k_B TV} = \rho_c \left[ \ln\left( \frac{\langle y_R(1_m)\rangle_s}{\langle y_R(1_m)y_R(2_m)\rangle_d} \frac{\xi-1}{4\Gamma} \right) + \ln\left( \frac{\Lambda^{3m}}{d^{3(m-1)}\Xi_m} \right) \right]$$
$$- \frac{\langle y_R(1_m)\rangle_s^2}{\langle y_R(1_m)y_R(2_m)\rangle_d} \frac{\xi-1}{4\Gamma}\left(1 + \frac{\xi-1}{4}\right) - \rho_c \ln\frac{\lambda^{m-1}}{2} \tag{36}$$

The contribution containing λ contains the energy of the covalent bonds within the chain molecule. This term can be neglected as it is independent of temperature and does not contribute to the properties of the chain molecules.

$$\frac{A - A_R^{EX}}{k_B TV} = \rho_c \left[ \ln\left( \frac{\langle y_R(1_m)\rangle_s}{\langle y_R(1_m)y_R(2_m)\rangle_d} \frac{\xi-1}{4\Gamma} \right) + \ln\left( \frac{\Lambda^{3m}}{d^{3(m-1)}\Xi_m} \right) \right]$$
$$- \frac{\langle y_R(1_m)\rangle_s^2}{\langle y_R(1_m)y_R(2_m)\rangle_d} \frac{\xi-1}{4\Gamma}\left(1 + \frac{\xi-1}{4}\right) \tag{37}$$

Equation (37) can be simplified significantly if we make the following approximation

$$\langle y_R(1_m) y_R(2_m) \rangle_d \approx \langle y_R(1_m) \rangle_s^2 \qquad (38)$$

Employing (38) within the free energy Eq. (37) we obtain

$$\frac{A - A_R^{EX}}{k_B T N_c} = -\ln\langle y_R(1_m) \rangle_s + \ln\left(\frac{\Lambda^{3m}}{d^{3(m-1)}\Xi_m}\right) + \ln\left(\frac{\xi - 1}{4\Gamma}\right) - \frac{2}{\xi + 1}\left(1 + \frac{\xi - 1}{4}\right) \qquad (39)$$

Where $N_c$ is the number of chain molecules in the system and $\xi$ is simplified to

$$\xi = \sqrt{1 + 8\rho_c \Gamma} \qquad (40)$$

The first two terms on the right hand side are the contributions which would have been obtained if we had employed the single chain approximation and truncated Eq. (5) after $c_1^{(o)}$, while the third and fourth term provide corrections to the single chain approximation; including contributions for chain-chain interactions. Note, the ideal contribution is included on the right hand side of (39).

As a check, we take the ideal gas limit if Eq. (39) by letting $\langle y_R(1_m) \rangle_s = 1$ and expanding $\xi$ in a first order Taylor's series to obtain

$$\left(\frac{A}{k_B T N_c}\right)_{ig} = \ln\left(\rho_c \frac{\Lambda^{3m}}{d^{3(m-1)}\Xi_m}\right) - 1 \qquad (41)$$

Equation (41) has the expected density dependence for an ideal gas of chains. It is interesting to note how the theory corrected the ideal free energy with a density independent intramolecular contribution $-\ln \Xi$, which accounts for the intramolecular free energy of an isolated chain.

The chemical potential is calculated from the free energy Eq. (39) as (see appendix)

$$\frac{\mu - m\mu_R^{EX}}{k_B T} = -\ln\langle y_R(1_m) \rangle_s - \rho_c \frac{\partial \ln\langle y_R(1_m) \rangle_s}{\partial \rho_c} + \ln\left(\frac{\xi - 1}{4\Gamma}\right) \qquad (42)$$

Where $\mu_R^{EX}$ is the excess chemical potential of the reference system. Also in Eq. (42) we have neglected (for compactness) any density independent terms which do not contribute to phase equilibria. Similarly, the compressibility factor of chain molecules $Z_c = P/\rho_c k_B T$ is obtained as

$$Z_c = mZ_R^{EX} - \rho_c \frac{\partial \ln\langle y_R(1_m)\rangle_s}{\partial \rho_c} + \frac{2}{\xi+1}\left(1 + \frac{\xi-1}{4}\right) \tag{43}$$

Adding and subtracting 1 we obtain

$$Z_c = Z_s + \Delta Z_d \tag{44}$$

Where $Z_s$ is the contribution due to standard TPT with a single chain interacting with the reference fluid

$$Z_s = 1 + mZ_R^{EX} - \rho_c \frac{\partial \ln\langle y_R(1_m)\rangle_s}{\partial \rho_c} \tag{45}$$

and $\Delta Z_d$ is a correction to the TPT result due to chain – chain interactions

$$\Delta Z_d = \frac{1}{2}\frac{1-\xi}{1+\xi} \tag{46}$$

Equation (22) gives the second virial coefficient of chain molecules in terms of cluster integrals over Mayer functions. Equivalently, $B_{2c}$ is defined as the zero density limit of the density derivative of the compressibility factor Eq. (43)

$$B_{2c} = \left.\frac{\partial Z_c}{\partial \rho_c}\right|_{\rho_c=0} = m^2 B_R - \left.\frac{\partial \langle y_R(1_m)\rangle_s}{\partial \rho_c}\right|_{\rho_c=0} + \frac{1}{2}\frac{\partial}{\partial \rho_c}\left(\frac{1-\xi}{1+\xi}\right)_{\rho_c=0} \tag{47}$$

Evaluating the derivatives on the right hand side

$$B_{2c} = m^2 B_R - \left.\frac{\partial \langle y_R(1_m)\rangle_s}{\partial \rho_c}\right|_{\rho_c=0} - \Gamma \tag{48}$$

Comparing Equations (48) and (27) we solve for $C$ defined in Eq. (30)

$$C = -\frac{\partial \langle y_R(1_m)\rangle_s}{\partial \rho_c}\bigg|_{\rho_c=0} \tag{49}$$

With (49) the temperature dependent $\Gamma$ can now be expressed as

$$\Gamma = m^2 B_R - \frac{\partial \langle y_R(1_m)\rangle_s}{\partial \rho_c}\bigg|_{\rho_c=0} - B_{2c} \tag{50}$$

We have demonstrated above, that the dual chain perturbation theory reduces to the exact second virial equation of state for chains in the low density limit. The results derived here are applicable to a wide range of reference fluids Lennard – Jones, hard sphere, square well etc… In the following section we apply this approach to a fluid of hard chains.

**III: Athermal chains of hard spheres**

The general results of section II are applied to the case of athermal hard chains composed of tangentially bonded hard spheres of diameter $d$ with a reference fluid potential

$$\phi_R(r) = \phi_{HS}(r) = \begin{cases} 0 & r \geq d \\ \infty & otherwise \end{cases} \tag{51}$$

The hard sphere reference system excess compressibility factor is obtained from Carnahan and Starling[19]

$$Z_{HS}^{EX} = \frac{1+\eta+\eta^2-\eta^3}{(1-\eta)^3} - 1 \tag{52}$$

At this point it is convenient to introduce the notation TPT$N_s$ and TPT$N_d$. TPT$N_s$ is $N$'th order perturbation theory in the single chain approximation. This is the classic form of thermodynamic perturbation theory introduced by Wertheim. In first order ($N = 1$) each bond is treated independently and there are no intra-chain correlations beyond nearest neighbors. In TPT1$_s$ information on bond angle or chain stiffness cannot be included. At second order ($N = 2$), intramolecular correlations are introduced between second nearest neighbors in the chain. These additional correlations allow for the inclusion of bond angle and chain stiffness degrees of freedom. The maximum order of perturbation theory form a chain is $N = m - 1$. TPT$N_d$ stands for $N$'th order dual chain perturbation theory and is related to TPT$N_s$ through Eqns. (44) and (46).

In TPT1, the multi-body cavity correlation function of a chain of tangentially bonded hard spheres is approximated as the product pair functions between bonded pairs

$$\langle y_{HS}(1_m) \rangle_s = y_{HS}(d)^{m-1} \tag{53}$$

Where $y_{HS}(d)$ is the hard sphere cavity correlation function at contact given by the Carnahan and Starling result[19]

$$y_{HS}(d) = \frac{1 - \frac{\eta}{2}}{(1-\eta)^3} \tag{54}$$

The packing fraction $\eta = \pi d^3 m\rho_c/6$. With Eqns. (53) and (54) the compressibility factor is obtained from Eqns. (44) and (50)

$$Z_{TPT1_d} = 1 + mZ_R^{EX} - (m-1)\rho_c \frac{\partial \ln y_{HS}(d)}{\partial \rho_c} + \frac{1}{2}\frac{1-\xi}{1+\xi} \tag{55}$$

with $\Gamma$ given as

$$\Gamma = \Gamma_{TPT1} = \left(\frac{m^2}{4} + \frac{5m}{12}\right)\pi d^3 - B_{2c} \tag{56}$$

To include additional intra-chain correlations we go to second order perturbation TPT2, which includes density dependent contributions for repulsions between second nearest neighbors along the chain. To proceed, we follow a similar approach to that of Marshall and Chapman[20] and approximate the multi-body correlation function in terms of the TPT2 free energy[11] for chain formation

$$\langle y_{HS}(1_m) \rangle_s = y_{HS}(d)^{m-1} \frac{\left(1+\sqrt{1+4\gamma}\right)^m}{2^m \sqrt{1+4\gamma}} \tag{57}$$

Where $\gamma$ includes information on the triplet correlation function and was determined by Phan *et al.*[11] for a freely jointed chain of hard spheres as $\gamma = 0.2336\eta + 0.1067\eta^2$. With this second order correction we obtain

$$Z_{TPT2_d} = 1 + mZ_R^{EX} - (m-1)\rho_c \frac{\partial \ln y_{HS}(d)}{\partial \rho_c} - \frac{2\rho_c d\gamma/d\rho_c}{1+4\gamma}\left(m\frac{\sqrt{1+4\gamma}}{1+\sqrt{1+4\gamma}} - 1\right) + \frac{1}{2}\frac{1-\xi}{1+\xi} \tag{58}$$

With $\xi_{TPT2}$ obtained from Eqns. (50) and (57)

$$\Gamma = \Gamma_{TPT2} = \Gamma_{TPT1} - 0.03883\pi d^3 (m^2 - 2m) \tag{59}$$

The first 4 terms on the right hand side of Eq. (58) give the TPT2$_s$ compressibility factor of Phan *et al.*[11]

For $B_{2c}$ we use a simple power law fit to the Monte Carlo simulation results of Yethiraj *et al.*[21] for the second virial coefficient of freely jointed hard sphere ($m = 2 - 128$)

$$\frac{B_{2c}}{m^2 d^3} = 0.181087 + \frac{1.784788}{m^{0.57577}} + \frac{0.281306}{m^{2.59849}} \tag{60}$$

As can be seen in Figure 2, Eq. (60) correlates the data over the simulation range to a high degree of accuracy with an average error of 0.2%. In the polymer limit, Eq. (60) will not have the exact chain length scaling and other correlations maybe used (see ref[21] for a full discussion).

With Eq. (60) we have gathered all required pieces to calculate the compressibility factor of freely jointed athermal chains. Table 1 compares compressibility factors from both single and dual chain (first and second order) perturbation theories to the Monte Carlo simulation results of Dickman and Hall[22]. The general trend is that $TPT1_s$ overpredicts $Z_c$ with an average error between theory and simulation of 6.5 %. Adding additional intra-chain correlations, while still treating intermolecular repulsions at the hard sphere reference level, gives improved results with $TPT2_s$ which yields an average error of 3.9 %. Finally, accounting for connectivity effects in inter-chain repulsions through the dual chain contribution Eq. (46), $TPT2_d$ gives an average error of 2.9%. All results in this section are using a monomer reference fluid. As can be seen, adding intra-chain correlations and including chain – chain interactions increases the accuracy of the equation of state. Alternatively, it has been shown that using a dimer reference fluid[12] can significantly improve predictions of the equation of state. The last column of table 1 gives theoretical calculations using the dimer reference fluid at first order (TPT1-DIMER); as can be seen, predictions are very similar to those obtained by $TPT2_d$. It has been demonstrated previously[23,24], that TPT1-DIMER gives more accurate predictions than several integral equation theory approximations within Wertheim's multi-density formalism. One advantage of the dual chain theory (over the dimer reference) is that the dual chain theory gives the exact second virial coefficient.

Very recently, Zmpitas and Gross[25] applied thermodynamic perturbation theory at the $TPT3_s$ level. They found improved performance as compared to $TPT2_s$ as verified by comparison to new molecular simulation data for athermal chains. In Fig. 3 we compare low density prediction of this $TPT3_s$ theory to $TPT2_d$. As can be seen, the introduction of chain – chain interactions results in a larger increase in accuracy than inclusion of more intramolecular contributions at the single chain level. A point of future research will be to apply TPT3 at the dual chain level (TPT3$_d$).

**IV: Lennard – Jones reference**

Attractive interactions can be incorporated into the theory by adding an attractive perturbation to a hard chain reference fluid[8,18], or by using an attractive reference fluid such as a fluid of Lennard – Jones (LJ)[26–28] spheres which interact with the LJ potential in Eq. (61)

$$\phi_R(r) = \phi_{LJ}(r) = 4\varepsilon\left(\left(\frac{d}{r}\right)^{12} - \left(\frac{d}{r}\right)^{6}\right) \tag{61}$$

Employing an LJ reference seems particularly promising with the dual chain perturbation theory due to the fact that the incorporation of chain – chain interaction, means that the theory should be able (to some extent) to predict the effect of branching on intermolecular attractions between chains.

For a LJ reference fluid in first order dual chain perturbation theory the compressibility factor is given by

$$Z_{TPT1_d} = 1 + mZ_{LJ}^{EX} - (m-1)\rho_c \frac{\partial \ln y_{LJ}(d)}{\partial \rho_c} + \frac{1}{2}\frac{1-\xi}{1+\xi} \tag{62}$$

with $Z_{LJ}^{EX}$ obtained by the equation of state of Johnson *et al.*[29] and $y_{LJ}(d)$ is evaluated using the correlation of Johnson *et al.*[26]

$$y_{LJ}(d) = \sum_{i=1}^{5}\sum_{j=1}^{5} a(i,j)(\rho^*)^i (T^*)^{j-1} \tag{63}$$

Where $a(i,j)$ are empirical constants, $\rho^* = \rho d^3$ and $T^* = \varepsilon/k_bT$. Finally, $\Gamma$ is given by

$$\Gamma = m^2 B_{LJ} - m(m-1) \left. \frac{\partial y_{LJ}(d)}{\partial \rho^*} \right|_{\rho=0} - B_{2c} \tag{64}$$

Chiew and Sabesan[30] simulated the second virial coefficient of tangentially bonded LJ chains $B_{2c}$ at lengths $m = 2 - 48$ and reduced temperatures $T^* = 4$, 5 and 8.

When Eq. (64) is evaluated using (63), the discriminate in Eq. (40) becomes negative and the theory gives imaginary results. This highlights a point which must be considered when constructing theories from this approach. Tracing the full approximation path

$$\xi = \sqrt{1 + 8\rho_c \frac{\langle y_{LJ}(1_m) y_{LJ}(2_m) \rangle_d}{\langle y_{LJ}(1_m) \rangle_s^2} \Gamma} \approx \sqrt{1 + 8\rho_c \left( m^2 B_{LJ} - m \left. \frac{\partial \langle y_{LJ}(1_m) \rangle_s}{\partial \rho} \right|_{\rho=0} - B_{2c} \right)} \tag{65}$$

$$\approx \sqrt{1 + 8\rho_c \left( m^2 B_{LJ} - m(m-1) \left. \frac{\partial y_{LJ}(d)}{\partial \rho} \right|_{\rho=0} - B_{2c} \right)}$$

The first approximation is that the dual chain average of the cavity correlation functions is given as the product of single chain averages $\langle y_R(1_m) y_R(2_m) \rangle_d \approx \langle y_R(1_m) \rangle_s^2$, the second is that the multi-body cavity correlation function can be approximated with the linear superposition $\langle y_{HS}(1_m) \rangle_s = y_{HS}(d)^{m-1}$ and the final approximation is that $\gamma = (\partial y_{LJ}/\partial \rho)_{\rho=0}$ is evaluated with the correlation in Eq. (62). It is this final approximation which is easiest to fix, and also appears to be the source of the problem. The term $\Gamma$ contains a difference between two quantities which are known "exactly" $B_{2c}$ and $B_{LJ}$ and a quantity which we are approximating $\gamma = (\partial y_{LJ}/\partial \rho)_{\rho=0}$. It is the inconsistency between the "exact" and the approximate that breaks the theory. To recover order we simply evaluate γ using the diagrammatic definition of the pair cavity correlation function to obtain the exact derivative

$$\gamma = \left.\frac{\partial y_{LJ}(d)}{\partial \rho}\right|_{\rho=0} = \int_{|\vec{r}_{13}|=d} d\vec{r}_{12} f_{LJ}(\vec{r}_{12}) f_{LJ}(\vec{r}_{23}) \tag{66}$$

The integration in Eq. (66) is restricted to the surface $|\vec{r}_3 - \vec{r}_1| = d$. Equation (66) can be easily evaluated numerically.

Table 2 list all relevant terms to evaluate the dual chain contributions Γ (40) and ultimately ξ (50) for two isotherms at $T^* = 4$ and 5 for chains ranging in size from $m = 2 – 48$. Using the exact γ evaluated by Eq. (66), the theory is well behaved as can be seen by the calculation of ξ at both low (η = 0.1) and high (η = 0.47) densities. Table 3 compares theory predictions of the chain compressibility factor for chains of LJ spheres to the molecular dynamics simulation results of Johnson et al.[26] for the isotherm $T^* = 4$. Similar to the hard sphere reference case, TPT1$_s$ tends to predict compressibility factors which are too large. Adding the dual chain correction gives improved agreement with simulation. The difference between TPT1$_s$ and TPT1$_d$ is most significant at lower densities.

**V: Non – linear molecules and associating fluids**

Throughout this document we have restricted discussion to linear chain molecules. Extension to non – linear molecules such as branched chains and rings is obtained by replacing the pre – factors of Eqns. (7) and (10) by ½ → 1/ν and ¼ → 1/ν² respectively, where ν is the symmetry number. Following through the derivation, Eq. (37) remains unchanged. What will change is the second virial coefficient and the appropriate approximation of the averaged correlation function Eq. (19).

As discussed in section II, the derivation in this paper is restricted to the limit of complete chain formation where smaller chain graphs will be negligible. To extend this formalism to two site chain forming associating fluids (hydrogen fluoride for instance), one would need to sum over all possible chain lengths. In the dual chain formalism this would require the second virial coefficient between chains of varying lengths. This is a tractable problem. Extending this approach to network forming fluids (more than two association sites) would require the enumeration of every possible associated cluster. This is not a tractable problem in the two density formalism; for systems such as these one would need to move to the multi – density $(>2)^4$ formalism. If one were able to reconstruct a similar dual chain theory in the multi – density formalism, for network forming fluids it would require all possible second virial coefficients between all possible associated clusters. This would likely prove prohibitively complex.

**V: Conclusions**

A new equation of state for chain molecules has been developed in the two density formalism of Wertheim. To guarantee the correct low density limit, it was necessary to move beyond the standard perturbation theories, and include contributions for chain - chain interactions. The theory was shown to reduce exactly to the second viral equation of state at low density. The theory was applied to both hard sphere and Lennard – Jones reference fluids and was shown to give improved results as compared to existing perturbation theories. For the LJ reference fluid, it was illustrated how a mix of approximate and exact input to the theory can give erroneous results. Care must be taken when applying this approach to new types of reference fluids to ensure the discriminate in Eq. (65) is non – negative.

**Appendix: Calculation of chemical potential**

The chemical potential is calculated from the free energy Eq. (39) as

$$\frac{\mu - m\mu_R^{EX}}{k_B T} = \frac{\partial}{\partial \rho_c}\left(\rho_c \frac{A - A_R^{EX}}{k_B T N_c}\right) = -\ln\langle y_R(1_m)\rangle_s - \rho_c \frac{\partial \ln\langle y_R(1_m)\rangle_s}{\partial \rho_c} + \ln\left(\frac{\xi-1}{4\Gamma}\right)$$
$$-\frac{2}{\xi+1}\left(1+\frac{\xi-1}{4}\right) + \Psi \quad (A1)$$

with

$$\Psi = \rho_c \frac{\partial}{\partial \rho_c}\left(\ln\left(\frac{\xi-1}{4\Gamma}\right) - \frac{2}{\xi+1}\left(1+\frac{\xi-1}{4}\right)\right) = \rho_c \frac{\partial \xi}{\partial \rho_c}\left(\frac{1}{\xi-1} + \frac{2}{(\xi+1)^2}\left(1+\frac{\xi-1}{4}\right) - \frac{1}{2(\xi+1)}\right) \quad (A2)$$

From Eq. (40)

$$4\Gamma\rho_c = \frac{\xi^2 - 1}{2} = \frac{(\xi-1)(\xi+1)}{2} \quad (A3)$$

and

$$\rho_c \frac{\partial \xi}{\partial \rho_c} = \frac{4\Gamma\rho_c}{\xi} = \frac{\xi^2 - 1}{2\xi} \quad (A4)$$

Then

$$\Psi = \frac{1}{2\xi}\left(\xi + 1 + \frac{2(\xi-1)}{\xi+1}\left(1+\frac{\xi-1}{4}\right) - \frac{1}{2}(\xi-1)\right) = \frac{2}{\xi+1}\left(1+\frac{\xi-1}{4}\right) \quad (A5)$$

Combining (A1) and (A5)

$$\frac{\mu - m\mu_R^{EX}}{k_B T} = -\ln\langle y_R(1_m)\rangle_s - \rho_c \frac{\partial \ln\langle y_R(1_m)\rangle_s}{\partial \rho_c} + \ln\left(\frac{\xi-1}{4\Gamma}\right) \quad (A6)$$

**References:**

1. Pratt, L. R. & Chandler, D. Interaction site cluster series for the Helmholtz free energy and variational principle for chemical equilibria and intramolecular structures. *J. Chem. Phys.* **66,** 147 (1977).
2. Morita, T. & Hiroike, K. A New Approach to the Theory of Classical Fluids. III. *Prog. Theor. Phys.* **25,** 537–578 (1961).
3. Wertheim, M. S. Fluids with highly directional attractive forces. II. Thermodynamic perturbation theory and integral equations. *J. Stat. Phys.* **35,** 35–47 (1984).
4. Wertheim, M. S. Fluids with highly directional attractive forces. III. Multiple attraction sites. *J. Stat. Phys.* **42,** 459–476 (1986).
5. Chapman, W. G. *PhD Disertation*. (1988).
6. Wertheim, M. S. Thermodynamic perturbation theory of polymerization. *J. Chem. Phys.* **87,** 7323 (1987).
7. Gil-Villegas, A. *et al.* Statistical associating fluid theory for chain molecules with attractive potentials of variable range. *J. Chem. Phys.* **106,** 4168 (1997).
8. Gross, J. & Sadowski, G. Perturbed-Chain SAFT: An Equation of State Based on a Perturbation Theory for Chain Molecules. *Ind. Eng. Chem. Res.* **40,** 1244–1260 (2001).
9. Llovell, F. & Vega, L. F. Prediction of thermodynamic derivative properties of pure fluids through the Soft-SAFT equation of state. *J. Phys. Chem. B* **110,** 11427–37 (2006).
10. Chapman, W. G., Gubbins, K. E., Jackson, G. & Radosz, M. New reference equation of state for associating liquids. *Ind. Eng. Chem. Res.* **29,** 1709–1721 (1990).
11. Phan, S., Kierlik, E., Rosinberg, M. L., Yu, H. & Stell, G. Equations of state for hard chain molecules. *J. Chem. Phys.* **99,** 5326 (1993).
12. Ghonasgi, D. & Chapman, W. G. A new equation of state for hard chain molecules. *J. Chem. Phys.* **100,** 6633 (1994).
13. Marshall, B. D. & Chapman, W. G. Three new branched chain equations of state based on Wertheim's perturbation theory. *J. Chem. Phys.* **138,** 174109 (2013).
14. Kalyuzhnyi, Y. V., Lin, C.-T. & Stell, G. Primitive models of chemical association. II. Polymerization into flexible chain molecules of prescribed length. *J. Chem. Phys.* **106,** 1940 (1997).
15. Kalyuzhnyi, Y. V., Lin, C.-T. & Stell, G. Primitive models of chemical association. IV. Polymer Percus–Yevick ideal-chain approximation for heteronuclear hard-sphere chain fluids. *J. Chem. Phys.* **108,** 6525 (1998).
16. Wertheim, M. S. Fluids with highly directional attractive forces. I. Statistical thermodynamics. *J. Stat. Phys.* **35,** 19–34 (1984).
17. Kierlik, E. & Rosinberg, M. L. A perturbation density functional theory for polyatomic fluids. II. Flexible molecules. *J. Chem. Phys.* **99,** 3950 (1993).
24

18. Chapman, W. G., Gubbins, K. E., Jackson, G. & Radosz, M. SAFT: Equation-of-state solution model for associating fluids. *Fluid Phase Equilib.* **52,** 31–38 (1989).
19. Carnahan, N. F. Equation of State for Nonattracting Rigid Spheres. *J. Chem. Phys.* **51,** 635 (1969).
20. Marshall, B. D. & Chapman, W. G. Molecular theory for self assembling mixtures of patchy colloids and colloids with spherically symmetric attractions: the single patch case. *J. Chem. Phys.* **139,** 104904 (2013).
21. Yethiraj, A., Honnell, K. G. & Hall, C. K. Monte Carlo calculation of the osmotic second virial coefficient of off-lattice athermal polymers. *Macromolecules* **25,** 3979–3983 (1992).
22. Dickman, R. & Hall, C. K. High density Monte Carlo simulations of chain molecules: Bulk equation of state and density profile near walls. *J. Chem. Phys.* **89,** 3168 (1988).
23. Stell, G., Lin, C.-T. & Kalyuzhnyi, Y. V. Equations of state of freely jointed hard-sphere chain fluids: Theory. *J. Chem. Phys.* **110,** 5444 (1999).
24. Stell, G., Lin, C.-T. & Kalyuzhnyi, Y. V. Equations of state of freely jointed hard-sphere chain fluids: Numerical results. *J. Chem. Phys.* **110,** 5458 (1999).
25. Zmpitas, W. & Gross, J. A new equation of state for linear hard chains:Analysis of a third-order expansion of Wertheim's Thermodynamic Perturbation Theory. *Fluid Phase Equilib.* (2015). doi:10.1016/j.fluid.2015.11.017
26. Johnson, J. K., Mueller, E. A. & Gubbins, K. E. Equation of State for Lennard-Jones Chains. *J. Phys. Chem.* **98,** 6413–6419 (1994).
27. Chapman, W. G. Prediction of the thermodynamic properties of associating Lennard-Jones fluids: Theory and simulation. *J. Chem. Phys.* **93,** 4299 (1990).
28. Pàmies, J. C. & Vega, L. F. Vapor−Liquid Equilibria and Critical Behavior of Heavy n-Alkanes Using Transferable Parameters from the Soft-SAFT Equation of State. *Ind. Eng. Chem. Res.* **40,** 2532–2543 (2001).
29. Johnson, J. K., Zollweg, J. A. & Gubbins, K. E. The Lennard-Jones equation of state revisited. *Mol. Phys.* **78,** 591–618 (1993).
30. Chiew, Y. . & Sabesan, V. Second virial coefficients of Lennard–Jones chains. *Fluid Phase Equilib.* **155,** 75–83 (1999).

**Tables**

| m | $\eta$ | $Z_c$ (MC) | TPT1$_s$ | TPT1$_d$ | TPT2$_s$ | TPT2$_d$ | TPT1-DIMER |
|---|---|---|---|---|---|---|---|
| 4 | 0.1072 | 2.25 ± 0.06 | 2.370 | 2.298 | 2.318 | 2.277 | 2.257 |
| 4 | 0.205 | 4.73 ± 0.05 | 4.881 | 4.768 | 4.778 | 4.709 | 4.727 |
| 4 | 0.252 | 6.4 ± 0.17 | 6.821 | 6.693 | 6.693 | 6.612 | 6.658 |
| 4 | 0.262 | 7.46 ± 0.16 | 7.320 | 7.189 | 7.187 | 7.104 | 7.155 |
| 4 | 0.278 | 8.02 ± 0.11 | 8.194 | 8.059 | 8.052 | 7.966 | 8.024 |
| 4 | 0.289 | 8.7 ± 0.07 | 8.854 | 8.716 | 8.706 | 8.618 | 8.680 |
| 4 | 0.31 | 9.8 ± 0.14 | 10.265 | 10.121 | 10.106 | 10.013 | 10.080 |
| 4 | 0.323 | 10.93 ± 0.13 | 11.249 | 11.102 | 11.083 | 10.987 | 11.056 |
| 4 | 0.34 | 12.2 ± 0.1 | 12.683 | 12.532 | 12.507 | 12.409 | 12.475 |
| 4 | 0.359 | 13.5 ± 0.1 | 14.510 | 14.355 | 14.324 | 14.221 | 14.279 |
| 4 | 0.376 | 16.1 ± 0.17 | 16.377 | 16.217 | 16.181 | 16.075 | 16.118 |
| 4 | 0.399 | 18.7 ± 0.3 | 19.314 | 19.150 | 19.106 | 18.996 | 19.008 |
| 4 | 0.417 | 21.7 ± 0.4 | 22.003 | 21.835 | 21.785 | 21.672 | 21.650 |
| 4 | 0.437 | 25.1 ± 0.3 | 25.473 | 25.301 | 25.244 | 25.127 | 25.054 |
| 8 | 0.0659 | 1.9 ± 0.06 | 2.259 | 2.131 | 2.165 | 2.074 | 2.019 |
| 8 | 0.1306 | 3.79 ± 0.08 | 4.279 | 4.095 | 4.090 | 3.950 | 3.897 |
| 8 | 0.1765 | 5.84 ± 0.07 | 6.398 | 6.188 | 6.141 | 5.976 | 5.960 |
| 8 | 0.227 | 9.05 ± 0.23 | 9.671 | 9.440 | 9.338 | 9.152 | 9.197 |
| 8 | 0.267 | 12.43 ± 0.17 | 13.227 | 12.983 | 12.834 | 12.634 | 12.727 |
| 8 | 0.308 | 17.5 ± 0.3 | 18.090 | 17.833 | 17.634 | 17.422 | 17.539 |
| 8 | 0.332 | 21.9 ± 0.4 | 21.681 | 21.419 | 21.189 | 20.971 | 21.078 |
| 16 | 0.0802 | 3.76 ± 0.21 | 4.034 | 3.803 | 3.768 | 3.571 | 3.385 |
| 16 | 0.148 | 7.32 ± 0.23 | 8.564 | 8.282 | 8.066 | 7.818 | 7.615 |
| 16 | 0.2045 | 13.2 ± 0.4 | 14.481 | 14.174 | 13.789 | 13.514 | 13.408 |
| 16 | 0.231 | 15.9 ± 0.3 | 18.196 | 17.881 | 17.412 | 17.128 | 17.085 |
| 16 | 0.247 | 18.2 ± 0.3 | 20.805 | 20.485 | 19.966 | 19.676 | 19.671 |
| | **Average error** | | 6.5% | 4.5% | 3.9% | 2.9% | 2.8% |

**Table 1:** Comparison of theoretical and Monte Carlo (MC) simulation results[22] for the compressibility factor of freely jointed hard sphere chains. The last column gives theoretical predictions using the dimer[12] reference fluid theory.

| m | T* | $B_{2c}/d^3$ | γ | $B_{LJ}/d^3$ | Γ | ξ(η=0.1) | ξ(η=0.47) |
|---|---|---|---|---|---|---|---|
| 2  | 4 | 0.098   | 0.047 | 0.12 | 0.30   | 1.11 | 1.44 |
| 4  | 4 | -1.166  | 0.047 | 0.12 | 2.55   | 1.41 | 2.36 |
| 8  | 4 | -6.411  | 0.047 | 0.12 | 11.58  | 1.79 | 3.38 |
| 16 | 4 | -28.63  | 0.047 | 0.12 | 48.56  | 2.37 | 4.77 |
| 24 | 4 | -64.77  | 0.047 | 0.12 | 109.05 | 2.82 | 5.80 |
| 32 | 4 | -120.52 | 0.047 | 0.12 | 198.73 | 3.24 | 6.75 |
| 40 | 4 | -201.3  | 0.047 | 0.12 | 323.04 | 3.65 | 7.68 |
| 48 | 4 | -290.49 | 0.047 | 0.12 | 465.34 | 3.98 | 8.40 |
| 2  | 5 | 1.05    | 0.058 | 0.26 | -0.14  | 0.94 | 0.70 |
| 4  | 5 | 2.15    | 0.058 | 0.26 | 1.25   | 1.21 | 1.80 |
| 8  | 5 | 4.65    | 0.058 | 0.26 | 8.47   | 1.62 | 2.93 |
| 32 | 5 | 46.48   | 0.058 | 0.26 | 157.90 | 2.92 | 6.04 |
| 48 | 5 | 103     | 0.058 | 0.26 | 355.45 | 3.51 | 7.36 |

**Table 2:** Quantities used to evaluate Γ in Eq. (64) for a LJ reference fluid. . $B_{2c}$ are simulation results of Chiew and Sabeson[30] and γ was evaluated numerically through Eq. (66).

| $m$ | $\rho^*$ | $Z_c$ (MD) | TPT1$_s$ | TPT1$_d$ |
|---|---|---|---|---|
| 4 | 0.9 | 13.9±0.08 | 14.440 | 14.238 |
| 4 | 0.5 | 2.42±0.08 | 2.596 | 2.443 |
| 4 | 0.1 | 0.98±0.03 | 1.013 | 0.961 |
| 8 | 0.9 | 25.3±0.1 | 26.937 | 26.665 |
| 8 | 0.5 | 3.20±0.08 | 3.716 | 3.493 |
| 8 | 0.1 | 0.88±0.04 | 0.972 | 0.877 |
| 16 | 0.9 | 48.4±0.3 | 51.929 | 51.602 |
| 16 | 0.5 | 4.7±0.2 | 5.955 | 5.671 |
| 16 | 0.1 | 0.80±0.08 | 0.892 | 0.742 |
| **average error** | | | 10.2% | 6.0% |

**Table 3:** Comparison of model predictions and molecular dynamics simulations for the compressibility factors of tangentially bonded Lennard – Jones spheres at a reduced temperature of $T^* = 4$. The simulation results $Z_c$(MD) were calculated from the results of Johnson *et al.*[26]

**Figure 1:**

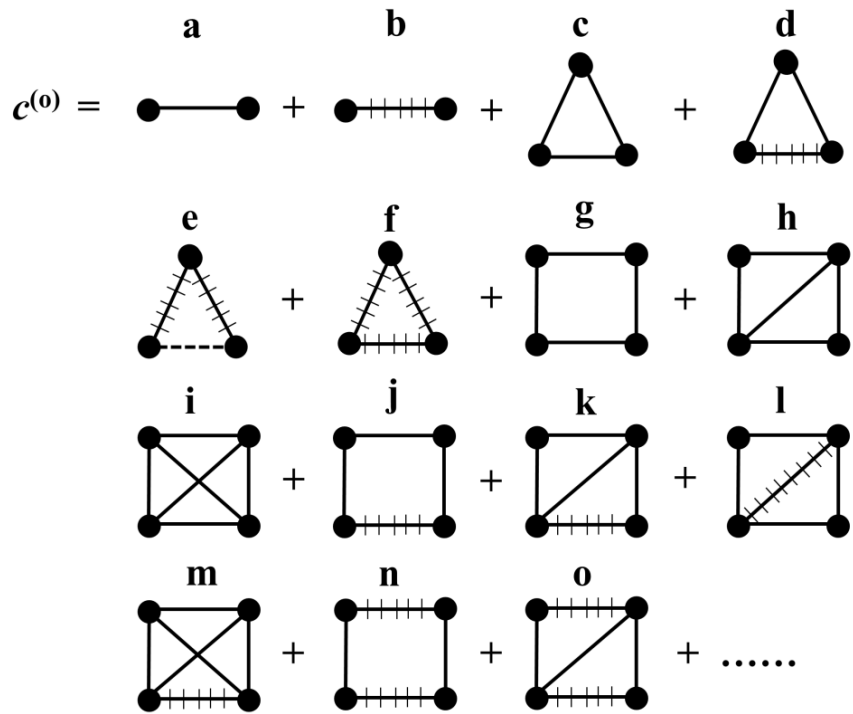

**Figure 1:** Graphical representation of Eq. (3) for dimerization where crossed lines ⊞ represent $F_A$ bonds, dashed lines represent $e_R$ bonds and solid lines represent $f_R$ bonds. Points with an incident $F_A$ bond are assigned monomer densities $\rho_o$, with all other points represented by the total density $\rho$. See the original publication[16] for more detail.

**Figure 2:**

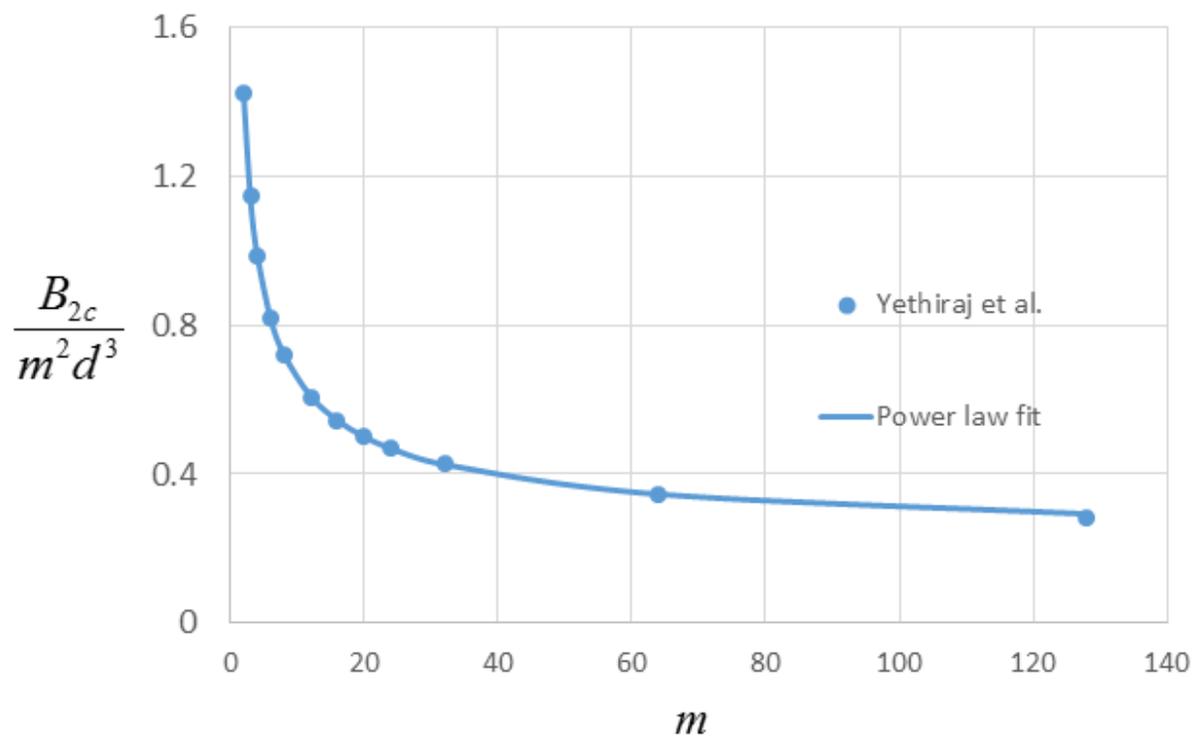

**Figure 2:** Fit of Eq. 60 to Monte Carlo simulations[21] of freely jointed hard chain second virial coefficients

**Figure 3:**

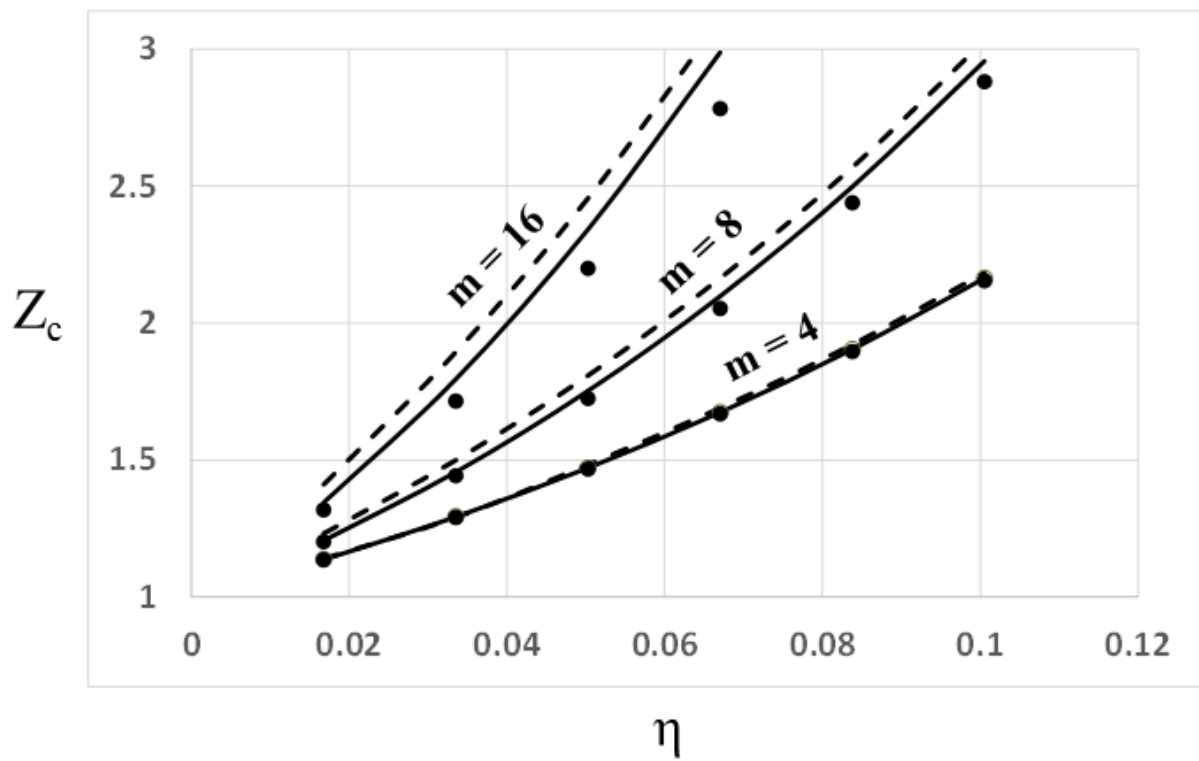

**Figure 3:** Comparison of TPT2$_d$ predictions (solid lines) to the TPT3$_s$ predictions (dashed curve) and Monte Carlo simulations (circles) of Zmpitas and Gross[25].